\documentclass[twocolumn,showpacs,preprintnumbers]{revtex4}

\usepackage{graphicx}
\usepackage{dcolumn}
\usepackage{bm}
\usepackage{amsmath,amssymb}

\newcommand{\n}{\normalfont~}

\newcommand{\bu}{B(E2)\negthinspace\protect\raisebox{0.03cm}{$\uparrow$} }
\newcommand{\parbu}{(B(E2)\negthinspace\protect\raisebox{0.03cm}{$\uparrow$}) }
\newcommand{\bua}{B(E2)\negthinspace\protect\raisebox{0.03cm}{$\uparrow$}$_{approx}$ }

\begin{document}

\preprint{}

\title{Probing the shell valence structure underlying the \bu for N,Z$\approx$40:\\
preponderance of the p-n interaction over the sub-shell closures.}

\author{I. Deloncle}
\email{deloncle@csnsm.in2p3.fr}
\homepage{http://www-csnsm.in2p3.fr/groupes/efix}
\affiliation{CSNSM, IN2P3/CNRS and Universit\'e Paris-Sud, F-91405
Orsay Campus, France}
\author{B. Roussi\`ere}
\email{roussier@ipno.in2p3.fr}
\affiliation{IPN, IN2P3/CNRS and Universit\'e Paris-Sud, F-91406 Orsay Cedex,France}

\date{\today}

\begin{abstract} The very simple product of
the number of particles by the number of holes appearing in the
expression of the reduced B(E2:\thickspace$0^+$$\rightarrow2^+_1$)
\parbu transition probability of even-even nuclei obtained from
the extension of the seniority scheme  is used to analyze the
experimental \bu values in the Cr up to Se isotopes. A new
interpretation is given to the \bu measured in $^{68}$Ni and
$^{70}$Zn. The \bu features of the even-even nuclei between Ni and
Se with neutron number ranging from 28 up to 50 fit in with a
global scenario involving p-n interaction. The evolution of the
\bu curves presenting very large values is amazingly reproduced by
very schematic binomial calculations.

\end{abstract}

\pacs{21.10.-k, 23.20.-g, 21.30.-x,
21.60.-n,27.40.+z,27.50.+e,27.60.+j}
\maketitle

\section{\label{sec:intro}Introduction\protect}

Quite recently new experimental results have been obtained which
can contribute to answer the questions concerning the $N,Z\simeq
40$ sub-shell closures. Coulomb excitation experiments performed
at Ganil have led to the measurement of the
B(E2:\thickspace$0^+$$\rightarrow2^+_1$) \parbu reduced transition
probabilities for the first time in the neutron rich $^{68}$Ni
\cite{Sor02} and $^{70}$Zn nuclei \cite{Sor01}. These data put
light on a different behavior of the \bu values in the Ni
isotopes, for N between 38 and 40, as compared to the
corresponding Zn isotopes: the \bu value decreases in the Ni
isotopes whilst it increases in the Zn ones. This has been
interpreted \cite{Sor01} putting forward, firstly, the change in
parity, which is brought by the $\nu 1g9/2$ orbital, hindering the
quadrupole E2 excitations (they preserve the parity symmetry), and
 leading to the strong reduction of the \bu
value of the spherical $^{68}$Ni \cite{Grawe,Sor02}(the $\nu
1g9/2$ are the levels above N=40). Secondly, has been put forward
a greater deformation in the $^{70}$Zn isotope which allows to
escape from $\nu 1g9/2$ by giving access to a great amount of
other orbitals \cite{Sor01}. But, it has been also mentioned that
in $^{68}$Ni neutron pair scattering plays an important role in
the \bu value \cite{Sor02}. Such pair excitations ((2n)p-(2n)h
type) cannot be affected by the parity of the $\nu 1g9/2$ levels.
In a recent publication \cite{Lang03}, the
 value of the \bu lower in $^{68}$Ni than in the doubly-magic
$^{56}$Ni nuclei is analyzed and explained, in $^{68}$Ni, by the
neutron pair scattering across the N=40 gap.\\Therefore, when
looking to the \bu curves in the nuclei of this region, the
decrease of the \bu between the $^{66}$Ni and$^{68}$Ni represents
a difference not only with the Zn isotopes, but with all their
neighbors. The \bu value between N=38 and 40 increases not only in
the Zn, but also in the Ge and in the Se isotopic chains, with in
addition for the \bu curves of the Zn and Se isotopes a minimum at
N=38.\\Nowadays, the region of the Ni isotopes is then the subject
of several studies. A new experiment on Coulomb excitation of the
neutron rich $^{70}$Ni and $^{72}$Zn isotopes has just been
performed at Ganil \cite{Sor02b}. Another experiment on $^{70}$Ni
will be performed at REX-ISOLDE \cite{Mayet02}. But already,
without these forthcoming results, the great amount of measured
\bu values makes possible systematical analyzes including nuclei
of the stability valley as well as many neutron rich nuclei. In
this paper, from such a work we propose a global interpretation of
the whole region surrounding $N,Z\simeq 40$, and new clues to
explain the difference between Ni and Zn isotopes.\\\indent The
\bu value is correlated to the possibilities to perform
excitations leading from the single-particle level scheme of the
$0^+$ to a $2^+$ state. So, very low \bu values are obtained for
nuclei with a closed shell since any excitation has to overcome
the gap. And, inversely, high values are reached at mid-shell
where the number of valence particles is maximal with respect to
the number of free levels, then available, for their excitations.
Consequently, in a very phenomenological approach, we propose to
analyze the \bu curves for nuclei having numbers of nucleons
ranging between 28 and 50 in terms of the underlying valence
space. In this approach, some of the \bu curves of these nuclei
appear to be consistent with a sub-shell structure inside the
28-50 major shell for both nucleon species, some with only one
sub-shell closure and some without any sub-shell closure. In the
Ni isotopes, due to the Z=28 gap, the \bu values reveals a complex
single-particle level scheme of the neutron valence space. The
part played by the p-n interaction in the onset and in the number
of sub-shell closures in the 28-50 major shells felt by the \bu is
discussed.

\section{\label{sec:senior}A simple formula for \bu}

The simplest formula relating the \bu transition probability to
the number of elementary excitations is provided by the seniority
scheme. In that scheme, assuming the N valence particles in a
single j-shell configuration (the single j-shell 1f$_{7/2}$ for
example), the E2 transition takes place between a $\nu$=0
seniority state (the N particles are all paired) and a $\nu$=2
seniority state (i.e. with one "broken" or aligned
pair)\cite{Talmi}. The \bu is then expressed as~:
\begin{equation}
\text{\bu} = \frac{1}{5}\times|\langle j^NJ=2^+_1\|Q\|j^NJ=0^+_1\rangle|^2\\
\end{equation}
and it can be shown \cite{RingSchuck} that~:
\begin{equation}
\text{\bu} \propto \text{\bua}
=\frac{N}{2}\times(\frac{\Omega_j}{2}-\frac{N}{2})
\end{equation}
with \(\Omega_j = (2j + 1) \).\\

\begin{figure}[h]
\vspace{-0.2cm}
\includegraphics[scale=0.5]{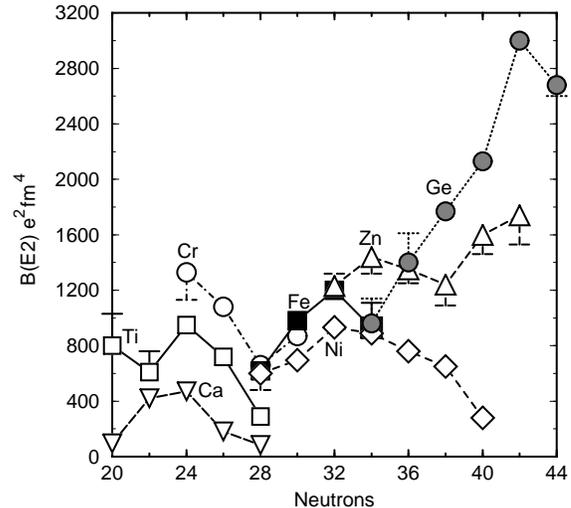}
\vspace{-0.3cm}
\caption{Joined with different lines and using different symbols,
experimental \bu values, for $N\geq20$ up to $N\leq44$, of Ca, Ti,
Cr, Fe, Ni, Zn  and Ge isotopes. All these values are taken from
\cite{Raman2}. For sake of legibility only a half of the error
bars, when larger than the symbols, is reported.}
\label{fig:fig1}
\end{figure}

The right hand side of (2) is nothing but the product of the
number of pairs of particles ($N_{part.}=\frac{N}{2}$) by the
number of pairs of holes
($N_{hol.}=\frac{\Omega_j}{2}-\frac{N}{2}$). This product gives,
as a function of N, a bell-shaped curve, symmetrical around the
maximum at N = $\Omega_j$/2, with two minima at N=0 and
N=$\Omega_j$. From such a bell-shaped curve one can thus deduce
$\Omega_j$, the beginning and the end of a single j-shell. The
bell-shaped profile of the \bu is also observed in nuclei for
which the valence shell extends over several single j-shells. The
\bu curve of the Ni isotopes, reported in fig.\ref{fig:fig1},
exhibits such a profile when, even under the assumption of a
sub-shell closure at N=40, the neutron valence space of the
$^{58-68}$Ni isotopes includes at least two single j-shells, the
1f$_{5/2}$ and the 2p$_{3/2}$ one. Moreover these different
j-shells seem, in first approximation, to have the same weights in
the \bu values. Indeed the bell-shaped profile for the Ni isotopes
appears even more regular than the one obtained for the Ca
isotopic chain, which is one of the textbook case for the (single
j-shell) seniority model \cite{Casten,Talmi}. Following this
constatation, and in order to treat the case of nuclei with
complex valence shell, we extend the single j-shell formula (2) by
replacing $\Omega_j$ by $\Omega$, sum of the $\Omega_j$ over the
different j-shells, and without, as in the quasi-spin formalism,
any weighting. We obtain for \bua in case of a complex valence
shell:
\begin{align}
\text{\bua} &= N_{part.} \times N_{hol.} \\
&= \frac{N}{2}\times (\frac{\Omega}{2}-\frac{N}{2})
\end{align}
\vskip -0.35cm
with \(\Omega = \sum \Omega_j\).\\[0.1cm]

Such an extension of the single j-shell case to a more complex
valence shell, through the quasi-spin formalism or the generalized
seniority one, is the basis of the boson models \cite{IBM79} which
allow a description of collective states. When the extension is
done through the generalized seniority, different weights are
assumed for the different j-shells. This appears to be essential
to reproduce the level spacing of the $^{58-66}$Ni isotopes
\cite{Talmi}. In this respect our approach can be found
over-simple. However, our aim is not to reproduce any values but
to analyze the \bu curves in a very phenomenological way using as
a reference the simplest model
calculations.\\
In that approach, the bell-shaped curve of the fig.\ref{fig:fig1}
can be understood as the sign of the conservation of the valence
space bulk structure (two gaps fixing the number of levels
confined between them) along an isotopic chain. The lower gap, in
particular, must be preserved even with a complete filling (up to
the upper gap) of the valence shell. Indeed, such a symmetrical
behavior implies that the values taken by $N_{part.}$ along a half
of the curve (with respect to the mid-shell) are taken
symmetrically by $N_{hol.}$ along the other half of the curve. The
number of levels must then be constant along the isotopic chain.
Indeed, as soon as the lower gap (the upper respectively) would
disappear, the levels of the shell which were below the gap (above
resp.) would enter in the valence space. The levels which were
below the lower gap being occupied, the number $N_{part.}$ would
be increased by the disappearance of the lower gap. The levels
which were above the upper gap being unoccupied, this is the
number $N_{hol.}$ which would be increased by the disappearance of
the upper gap. The disappearance of any of the two gaps
surrounding the shell, will then immediately and strongly induce
the increase of the product $N_{part.}\times
N_{hol.}$.\\Reciprocally, in such a model, any sharp discontinuity
of a \bu curve has to be interpreted as such a change in the
valence space, as observed at the end of a valence shell. In
fig.\ref{fig:fig1} the increase of the \bu curve of the Zn
isotopes from N=38 could be considered as such a discontinuity.
But it has been interpreted by the onset of deformation in
$^{70}$Zn \cite{Sor01} which can also change the number of levels
of the valence space. It stays to determine if a change of
deformation can lead to such a discontinuity. In this aim, in the
following section, we will study for all nuclei having N or Z
comprised between 8 and 52 (and some are well-known deformed
nuclei), the rate of the increase (or of the decrease) of the
experimental \bu values between two consecutive isotones (or
isotopes) by computing their ratio.

\section{\label{sec:rat}Ratios of experimental \bu values}

\begin{figure*}
\vspace{-1.5cm}
\hspace{-0.6cm}
\includegraphics[angle=-90,scale=0.70]{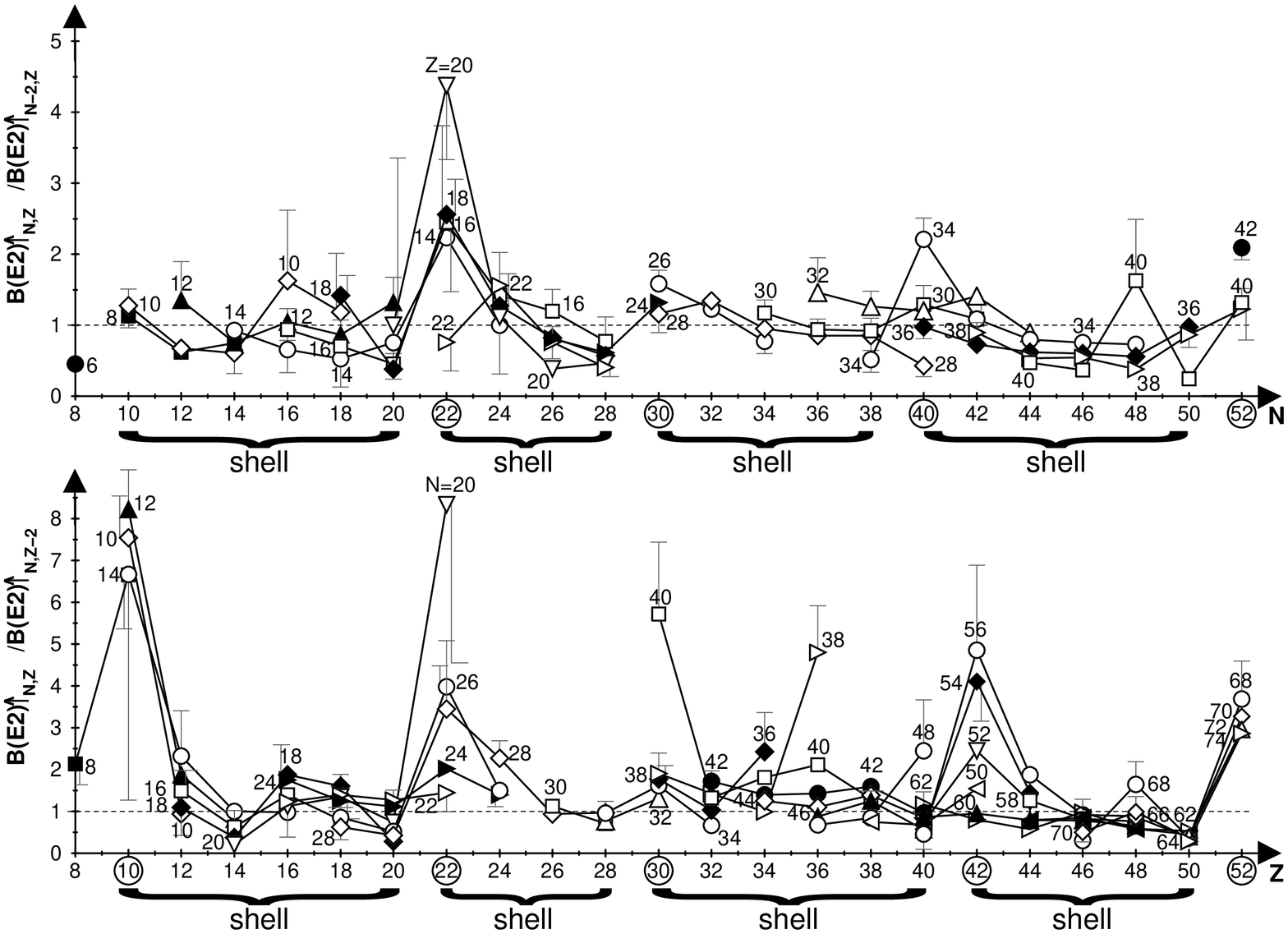}
\vspace{-0.5cm}
\caption[]{Upper part: For isotopic chains, Z ranging from 8 to 52,
ratios of measured B(E2)$\uparrow_{N}$/B(E2)$\uparrow_{N-2}$
values as a function of neutron number.\\Lower part: For isotonic
chains, N ranging from 8 to 52, ratios of measured
B(E2)$\uparrow_{Z}$/B(E2)$\uparrow_{Z-2}$ values as a function of
proton number. The \bu values ar taken from \cite{Raman2}. The
shells are explicitly located. For sake of legibility only a half
of the error bars, and when they are larger than the symbols, are
reported with grey lines.}
\label{fig:ratio}
\end{figure*}

In fig.\ref{fig:ratio} are reported, at given N, for the upper
part of the picture (and at given Z for the lower part), the
ratios of the experimental \bu values of two consecutive nuclei~:
B(E2)$\uparrow_{N(or Z))}$/B(E2)$\uparrow_{N-2 (or Z-2))}$.\\One
can note that~:\\
i) The behavior of the \bu, increasing up to the
maximum at mid-shell then decreasing within a bell-shaped profile,
induces parabola-like curves for the ratio of two consecutive \bu
values, starting from ratio values well above 1.0 ending with
values below 1.0 at the end of a shell.\\
ii) Due to the expression of the ratio, the shell closure located
at a given particle number N or Z has repercussions on the ratios
at N+2 (or Z+2) (i.e. at 22 for a shell closure at 20, 30 for a
shell closure at 28....).\\
iii) The (sub)-shell closures induce a mean increase and a
dispersion of the values of the ratios. This can be seen at 22,
30, 52. For the sub-shell closure around 40 this is also true, but
the effect is at 42 for the protons and 40 for the neutrons. The
sub-shell closure is then located by the ratios of experimental
\bu values at 38 in neutrons and 40 in protons.\\
iv) Different curves are obtained in a neutron (proton resp.)
shell according to the different Z (N resp.) values. This
indicates that the protons and the neutrons do not mainly
contribute to the \bu value by way of a product. In the next
sections we will take into account protons and neutrons by summing
their contributions.

In order to determine in what extent the increase of \bu value is
sensitive to the increase (or decrease) of deformation, we can
examine some of the curves of ratios corresponding to isotopic
series which are well known for their variation of deformation.
Isotope shift measurements \cite{Keim} have shown that the
deformations in the Kr (Z=36) and Sr (Z=38) isotopes are nearly
identical from N=38 up to 58. From N=40 up to 50, they decrease
from their maximal deformation, $\beta$=0.4, to a small
deformation, $\beta\sim$0.15 \cite{Keim}. This change of
deformation, large but progressive, has no particular effect on
the \bu ratios obtained for the [40-52] neutron valence shell and
reported in the upper part of fig.\ref{fig:ratio}. The curves for
the Kr and Sr isotopes are, as a function of N, regular compared
to what is observed for the [28-40] proton valence shell (lower
part of this figure) and the values taken along these curves are
comparable to the ones obtained for the [30-38] neutron valence
shell. One has to conclude that the changes of deformation in the
Kr as well as in the Sr isotopes do not lead to irregularities in
the evolution of the \bu values between two of their consecutive
isotopes. The increase (or decrease) of their \bu values, from
N=42 up to 48, are not exceeding the one observed between two
consecutive points in a bell-shaped \bu curve.

Another example is provided by the Ge isotopes the shape
instability of which is known \cite{Chuu93}~: transitions from a
weakly deformed (oblate \cite{Ardo75}) $^{68}$Ge$_{36}$, to the
nearly spherical $^{70}$Ge$_{38}$ \cite{Ardo75,Ardo78,Kumar78},
then back to a weakly deformed nucleus (oblate
\cite{Ardo75,Kumar78}) up to, at N=44 and 46, prolate deformation
according to \cite{Ardo75} or weakly oblate according to
\cite{Ardo78}. Moreover shape coexistence is a common phenomenon
in these isotopes, it has been studied in $^{68}$Ge$_{36}$
\cite{Lima81,Chat91,Herm92}, as well as in the
 $^{70,72,74}$Ge$_{38,40,42}$ \cite{Ronn76,Kumar78}.
Despite this rather complex situation, the curve of the \bu ratio
obtained for the Ge isotopes is very regular. In fact, for
N$\leq$38, the variation with N of the \bu ratios obtained for the
Ge, is very similar to the Zn and the Ni ones, and for N$\geq$40
up to 44, to the variation in the Kr and Sr isotopes. However,
below N=40, one can note that the distance between the Ge (Z=32)
curve and the Zn (Z=30) one, clearly greater than that between the
Zn and Ni (Z=28) curves --very close together--, is nearly
constant (or slightly decreasing). Such a difference, nearly
constant with N (at least without distortion of the curve), is
more probably the expression of a non linearity in the proton
contribution when going to the Ge isotopes, the same number of
protons separating the Ge from the Zn isotopes and the Zn from the
Ni isotopes. One can also note that above N=40, the distance
separating the curve of the Ge from the Se (Z=34) curve becomes
normal; these different distances will find an interpretation in
the next section. Concerning the relation between the
discontinuity of a \bu curve and a change in deformation, the Ge
isotopes lead to the same result as the Kr and Sr isotopes, their
changes in deformation do not induce irregularities in their \bu
ratios. Assuming a larger, and nearly constant, proton
contribution to the Ge \bu values, the increase (or decrease) of
these \bu values follow the one observed between two consecutive
points in a bell-shaped \bu curve.

At the opposite of these smooth behaviors, the curve of the \bu
ratios obtained for Z=40 (Zr isotopes) presents a strong
irregularity at N=48, such as, in the lower part of the
fig.\ref{fig:ratio}, the irregularity of the N=38 curve at Z=36.
The first case corresponds to the ratio of the $^{88}$Zr \bu value
over the $^{86}$Zr one, which is very low. The low-lying levels of
$^{86}$Zr have been interpreted as individual-particle excitations
of relatively few particles (with some vibrational collectivity)
\cite{Warburton85,Kaye98}. This interpretation is supported by
Hartree-Fock-Bogoliubov calculations predicting a spherical shape
for the vacuum configuration in this nucleus \cite{Kaye98}.
Unfortunately, no isotope shift experiments have been performed
for the Zr neutron deficient isotopes, and no direct (independent
from the \bu one) data are available for deformation of these two
nuclei. The second irregular point (at Z=36 in the N=38 curve of
the lower part of fig.\ref{fig:ratio}) involves the \bu value of
$^{74}$Kr as numerator and of $^{72}$Se as denominator. The
isotope shift measurements \cite{Keim} previously mentioned lead
to assign a large deformation ($\beta\gtrsim$0.4) to $^{74}$Kr. No
such measurements are available for the $^{72}$Se isotope, but
lifetime measurements lead to a deformation of $\beta\sim$0.18
\cite{Lieb77} for the ground state band which is interpreted as
vibrational up to I=6 \cite{Lieb77,Hees86}. Moreover this nucleus
is known to exhibit a shape coexistence \cite{Ham74} between the
ground state band and the low-lying excited one having
$\beta\sim$0.31. It is worth noting that this phenomenon has been
related to the competition between the shell gaps \cite{Chat91,
Chan00} appearing in the mean field at nucleon numbers 34, 36
(oblate), 34,38 (prolate), and 40 (spherical)
\cite{Ragn78,Naz85,Bonc85}. These gaps can induce the irregular
behavior, at Z=34,36 and 40, in the N=36-40 isotonic curves
reported in the lower part of fig.\ref{fig:ratio}. This irregular
behavior is not only characterized by the high values of the
ratios obtained at these Z values, but also in the non-
parallelism between the different curves in the Z=34-40 range as
compared to the different isotopic curves (upper part of the same
figure) between N=34 and 40. This behavior and the \bu values are
comparable to what is observed at the (sub)-shell closures. It is
worth recalling that in the Ge isotopes, the proton number (Z=32)
of which does not exactly correspond to one of the previous gaps,
the shape coexistence phenomenon, for example in $^{68}$Ge
\cite{Chat91}, does not induce any irregularity at N=36 in the
curve. It seems that these gaps are effective when both protons
and neutrons correspond to 34, 36, 38 or 40. Another particle
numbers which seem to induce such a reinforcement in
fig.\ref{fig:ratio} are 40, 48 (and 68).

In conclusion it appears that the ratio of the \bu of two
consecutive even-even isotopes (resp. isotones) is very sensitive
to the (sub)-shell closure, to neutron or proton gaps. Following,
the ratio can appear to depend on the deformation changes when
these changes are induced by gaps. From this result it seems
difficult to incriminate only an increase of deformation to
explain the discontinuity of the \bu curve at N=38 for the Zn
isotopes. The latter appear to highlight a N=38 gap that the Ni
isotopes do not feel. Between N=38 and 40, the evolution of the
\bu in the Ni isotopes, with the decrease for $^{68}$Ni, is then
stranger than the evolution in the Zn isotopes. Indeed, the \bu
value increases between N=38 and 40 not only in the Zn, but also
in the Ge and in the Se isotopes (at N=40 their \bu ratios are
greater than 1.0 as it can be seen in the upper part of the
fig.\ref{fig:ratio}).

The deformation being insufficient to explain this difference, we
will explore the track of the valence space change. Firstly, in
order to get other criteria than the increase or the decrease of a
\bu curve to sign a change of valence space, we will analyze the
results of calculations performed for various valence spaces
coming into play for nucleon numbers ranging from 20 up to 50.
This is presented in the next section. Then, we will use these
criteria to examine the experimental curves of nuclei having
neutron and proton numbers comprised between 20 and 50.


\section{\label{sec:calc}Results of calculations}

As mentioned in sec.\ref{sec:rat} we take into account both type
of particles by summing their contribution. The formula we use in
the calculations is then:
\begin{equation}
\text{\bua} =
[\frac{N_p}{2}\times(\frac{\Omega_p}{2}-\frac{N_p}{2})] +
[\frac{N_n}{2}\times(\frac{\Omega_n}{2}-\frac{N_n}{2})]
\end{equation}
where p stands for protons and n for neutrons.\\

For nucleon number ranging between 20 and 28, the only possible
valence space is the single-j shell $f_{7/2}$. For nucleon number
above 28, there are four possible combinations of proton and
neutron valence spaces:  for each type of nucleon, two cases of
valence space have to be considered, either the complete 28-50
major shell (this gives $\Omega_p=22$ for protons and
$\Omega_n=22$ for neutrons), either a sub-shell resulting from a
sub-shell closure located, accordingly to the observations of the
previous section, at N=38 for neutrons and at Z=40 for protons.
The proton sub-shell closure leads to two sub-shells, which have
$\Omega_p= 12$ for Z below 40 and $\Omega_p= 10$ for Z above, and
the neutrons sub-shell closure gives two neutron sub-shells, one
having $\Omega_n= 10$ for N below 38 and the other $\Omega_n= 12$
for N above.

Except the different localization of the sub-shell closure, the
results obtained as a function of N or as a function of Z lead to
the same criteria for pointing out a change in the valence space.
In this section we only present calculated \bua curves as a
function of N and, for sake of legibility, for only a few proton
number values. But at least, for each combination of proton and
neutron valence spaces, we present the curves of minimal and
maximal magnitude labelled by the proton number used for their
calculations.

\begin{figure}[h]
\vspace{-0.2cm}
\includegraphics*[scale=0.5]{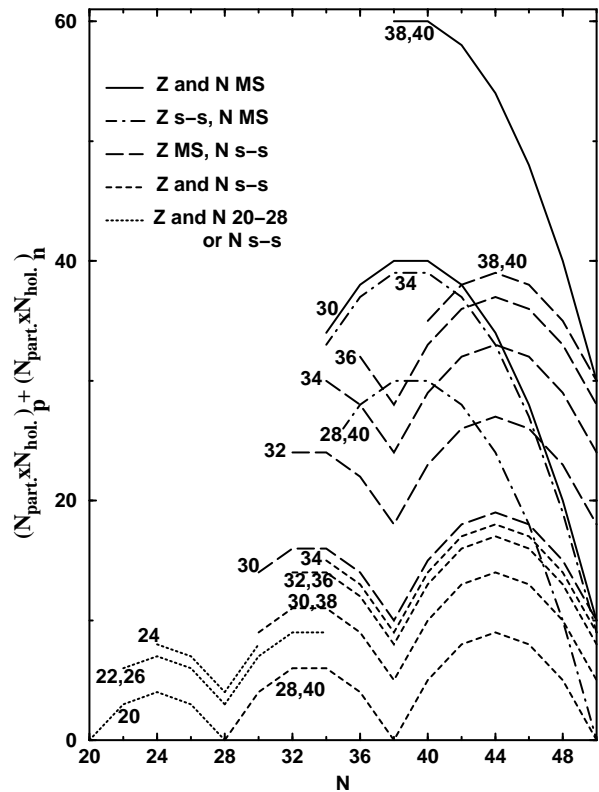}
\vspace{-0.3cm}
\caption[]{Calculated \bu curves for different
proton number assuming the following valence spaces: the major
shell for both type of nucleons (Z and N MS, solid lines); a
sub-shell closure in protons and the major shell in neutrons (Z
s-s, N MS, dot-dashed lines) ; the major shell in protons and a
sub-shell closure in neutrons (Z MS, N s-s, long-dashed lines);
sub-shell closure for both types of nucleons (Z and N s-s, in
short-dashed lines); single j-shell $f_{7/2}$ for protons and for
neutrons or a sub-shell ranging between 30-38 for neutrons (Z and
N 20-28 or N s-s, in dotted lines).}
\label{fig:results}
\end{figure}

In fig.\ref{fig:results}, for the lower N values
(20~$\leq$~N~$\leq$~32) the curves reported in dotted lines have
been calculated for proton number below 28. The corresponding
proton valence space is the single-j shell $f_{7/2}$ ($\Omega_p=
8$), but the neutron valence space is either the $f_{7/2}$ single
j-shell ($\Omega_n= 8$) when N$\leq$28, or a sub-shell ranging for
N$\geq$28 ($\Omega_n= 10$). These curves are specified by \bf{Z
and N 20-28 or N s-s}\n. For N$\geq$28 (and up to 50), the
calculations corresponding to the four cases of valence have been
performed, one has considered proton numbers from 28 up to 40 (in
the case of proton sub-shell closure, only the lower sub-shell
with $\Omega_p= 12$ comes into play).

In fig.\ref{fig:results} at N=38 from the bottom to the top one
finds the curves obtained assuming:\\
1) a proton sub-shell closure and a neutron sub-shell closure
(\bf{Z and N s-s}\n)\\
2) the 28-50 major shell in protons and a sub-shell closure in
neutrons (\bf{Z MS, N s-s}\n). The curve obtained for Z=28 is, in
the figure, merged with the Z=28 one of the \bf{Z and N s-s}\n curves.\\
3) a proton sub-shell closure and the neutron major shell (\bf{Z s-s, N MS}\n).\\
4) the major shell in protons and in neutrons ( \bf{Z and N
MS}\n). The curve obtained for Z=28 is, in the figure, merged with
the Z=28 one of the \bf{Z s-s, N MS}\n curves.\\

Several main features, related to the size of the neutron or of
the proton valence space, appear from the analysis of the
fig.\ref{fig:results}:\\
i) The presence of a sub-shell closure at N=38 induces a minimum
in the \bua curves, as do the shell-closures at 20, 28 and 50. In
case of a N=38 sub-shell closure one expects then an increase in
the N=38-44 range. On the contrary, the curves calculated with a
neutron major shell assumption, decrease in this range.\\
ii) The position of the maximum of each curve depends on the
neutron valence space. Since the maximal contribution of
$N_{part.} \times N_{hol.}$ is at mid-valence shell, it is at
N=38-40 in case of a major shell for neutrons valence, at N=24 for
a $f_{7/2}$ single-j shell, at N=32-34 for the lowest neutron
sub-shell resulting from a N=38 closure, and at N=44 for the
highest neutron sub-shell.\\
iii) The spacing between the curves and the amplitude, increase
with the size of the proton valence space. One can see the greater
spacing between the different curves drawn in long-dashed lines,
obtained assuming a major shell in protons and a sub-shell in
neutrons, as compared to the curves in short-dashed lines obtained
with a sub-shell in protons and a sub-shell in neutrons. On the
contrary, one observes identical spacings between the curves when
calculated with a same proton valence space. A same distance
separates the curve labelled 28,40 from the one labelled 34 in
short-dashed lines or in dot-dashed lines, all obtained with a
proton sub-shell but with a different neutrons valence space.
(This is easier to see at N=50 where the dot-dashed curves are
joined with the short-dashed ones.) There is also a same distance
between the curves labelled 30 and the ones labelled 38, 40 drawn
in long-dashed lines or in
solid lines, obtained with the proton major shell.\\
iv) The magnitude reached by the various curves depends on the
total valence space (protons plus neutrons), it increases with the
size of the total valence space~:  the curves obtained with a
major shell for both protons and neutrons (solid lines) reach the
highest magnitude, then one finds the curves corresponding to a
major shell for the valence space of one species and a sub-shell
for the other species (dot-dashed and long-dashed lines), and
finally comes the curves obtained with small valence space for
both species, sub-shell or the single j-shell $f_{7/2}$
(short-dashed and dotted lines). So, three different magnitudes
are reached in fig.\ref{fig:results}.\\
v) The order, in which the curves corresponding to different Z
values appear, depends only on the proton valence space. The
maximal proton contribution being obtained at the proton
mid-valence, the highest curve is obtained for Z=38 and 40 in
calculations performed with a proton major shell and at Z=34 for a
28-40 proton sub-shell. Moreover, and this will be used in the
following, one can note that in this latter case the Z=32,36
curves are under the Z=34 one.
\\Using these features as criteria determining the valence space size,
we propose, in the next section, to analyze the experimental \bu
curves.

\section{\label{sec:ana}Analysis of the experimental \bu curves}

\begin{figure}[h]
\vspace{-0.2cm}
\includegraphics[scale=0.5]{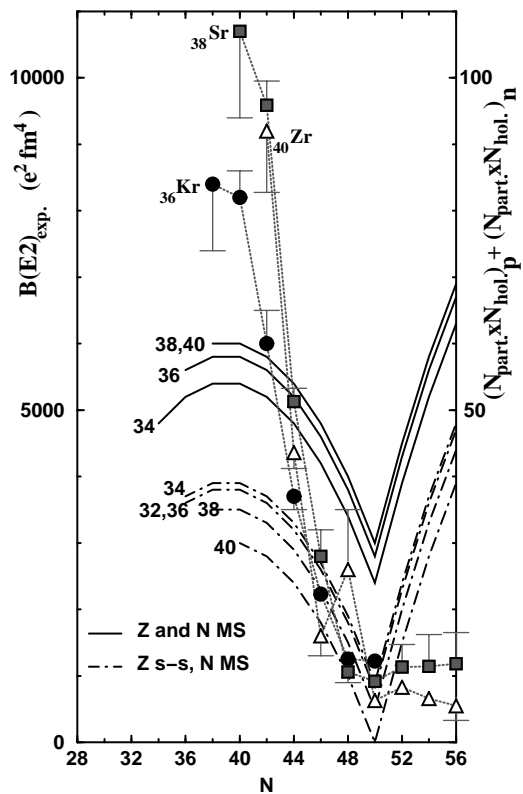}
\vspace{-0.3cm}
\caption[]{Joined by dotted grey lines and using different
symbols, experimental \bu values for the Kr , Sr and Zr isotopes
taken from \cite{Raman2}. For sake of legibility only a half of
the error bars (when larger than the symbols) is reported. Several
of the calculated \bua curves presented in fig.\ref{fig:results}
are reported for Z=34,36,38 and 40. For all the curves the proton
number is explicitly given.}

\label{fig:NpNha}
\end{figure}

\begin{figure}[h]
\hspace{-0.5cm}
\vspace{-0.2cm}
\includegraphics[scale=0.5]{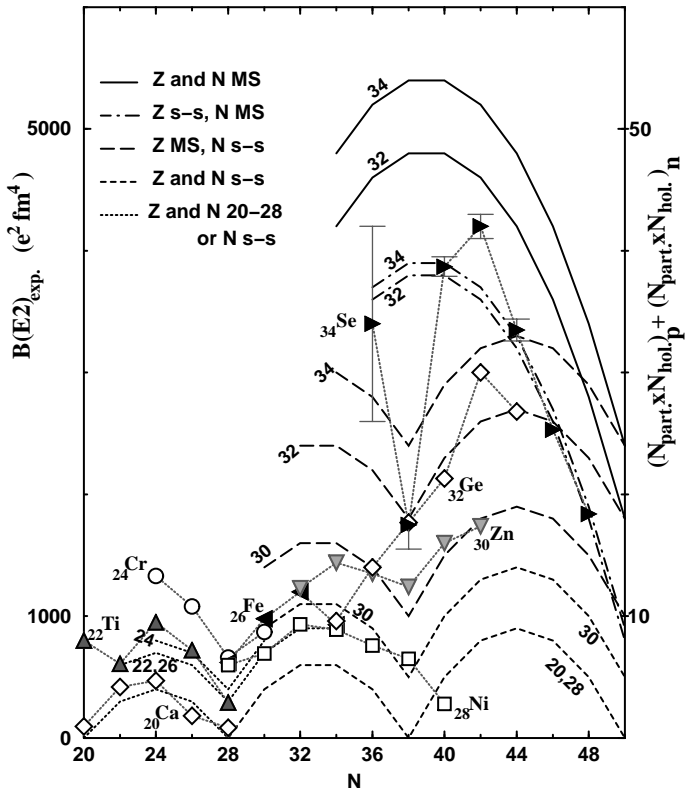}
\vspace{-0.3cm}

\caption[]{Joined by dotted grey lines and using different symbols,
experimental \bu values for Ca, Ti, Cr, Fe, Ni, Zn, Ge and Se
isotopes from \cite{Raman2}. For sake of legibility only a half of
the error bars of the Se isotopes (when larger than the symbols)
is reported. For the error bars on the other nuclei see
fig.\ref{fig:fig1}. Several of the calculated \bua curves
presented in fig.\ref{fig:results} are reported. For all the
curves the proton number is explicitly given.}

\label{fig:NpNhb}
\end{figure}

In fig.\ref{fig:NpNha} are reported as a function of N the
experimental \bu values for nuclei with Z=36, 38 and 40. In
fig.\ref{fig:NpNhb} are reported the \bu measured in nuclei with
20$\leq$Z$\leq$34. In both figures, the most relevant theoretical
curves of fig.\ref{fig:results} are reported using a scale giving
100 e$^2$fm$^4$ per unit of $(N_{part.} \times N_{hol.})_p +
(N_{part.} \times N_{hol.})_n$. This scaling factor (the value at
the origin is zero) has been chosen in order to reproduce the \bu
curve of the Ca isotopes for which only the single j-shell
$f_{7/2}$ plays a role in protons as well as in neutrons. Starting
from this, we can only discuss relative magnitudes or relative
amplitudes. In order to pass easier from one figure to the other
which have different vertical scales, we have drawn in
fig.\ref{fig:NpNha} and in fig.\ref{fig:NpNhb}, with solid and
dot-dashed lines, the two highest Z=34 calculated curves .

Before going further, let us make a remark on the relative
underestimation of some experimental values by the calculations in
fig.\ref{fig:NpNha}. The highest experimental values reported are
well above the highest calculated ones obtained assuming the
largest possible valence spaces for particle numbers around 40
(28-50 major shells). Nevertheless, for N$\geq$42, in particular
from N=44, the calculations have the good relative order of
magnitude as compared to the experimental values. In most cases,
the calculations are then in agreement with the experiment. In
fact, these features (good agreement for the low experimental
values and underestimation of the few highest values) reflect
mainly a large difference of the shapes between the calculated
curves and some of the experimental ones, and exclude a
systematical effect which could come from the seniority. The shape
difference will be detailed in the section.\ref{sec:proba}. The
tendency to the underestimation of the highest experimental values
of an isotopic chain by the calculations can be used as a criteria
to exclude some combinations of the proton and neutron valence
spaces~: those overestimating all the experimental values, or
those underestimating all the experimental values.

We will now proceed to the analysis of the figures using: the
order of magnitude, the position in N of the maxima, the existence
or not of a minimum at N=38, the order of the curves corresponding
to different Z values, the spacing between the curves, the
underestimation of some experimental values and the shape of the
curves.

As the calculated curves, the experimental \bu curves reported in
figs.\ref{fig:NpNha} and \ref{fig:NpNhb} present three orders of
magnitude~: the $_{36}$Kr, $_{38}$Sr and $_{40}$Zr isotopes
reported in fig.\ref{fig:NpNha} exhibit the highest values; the
Se, Ge, Zn isotopes reported in fig.\ref{fig:NpNhb} reach the
intermediate magnitude; the Ni, Fe, Ti and Ca isotopes lie at the
lowest magnitude.

In fig.\ref{fig:NpNha}, as obtained in calculations with a neutron
major shell, the experimental curves have a maximum below N=44~:
at N=38 for Kr, 40 or below for Sr and N$\leq$42 for Zr, from
which they decrease continuously up to N=50. On the contrary, in
fig.\ref{fig:NpNhb} experimentally there are two regions for a
maximum~: near N=42 for Se, Ge, and probably Zn; at N=32-34 for
Ni, 34 for Zn and probably Se. Between these two regions there is,
for nearly all the isotopic series, an increase between N=38-42
and even a minimum at N=38 in Se, Zn. The decrease between N=38-42
in the Ni isotopes appears to be nothing but only one exception.
Except this decrease, the latter features of fig.\ref{fig:NpNhb}
are in agreement with the characteristics of the calculated curves
assuming a neutron sub-shell closure. In the range N=20-28, the
experimental curves have a maximum at N=24 in Ca, Ti and probably
in Cr, in fact they follow pretty well the calculated curves.

In fig.\ref{fig:NpNha} for N$\leq$44, the order of the
experimental curves as the spacings between them, are similar to
the ones obtained with the assumption of a proton major-shell
(solid lines). The $_{38}$Sr and $_{40}$Zr \bu curves are very
close together and above the $_{36}$Kr one. All these curves lie
above the ones of fig.\ref{fig:NpNhb}, in particular above the
$_{34}$Se one. For N$\geq$46 in fig.\ref{fig:NpNha}, one can note
some inversions in the order of the experimental curves which
afterwards follow the order corresponding to a sub-shell closure
in protons (dot-dashed curves). In fig.\ref{fig:NpNhb}, the order
of the experimental Ca, Ti and Cr curves is in agrement with the
one obtain in the calculations. For the light nuclei of
fig.\ref{fig:NpNhb}, the calculated spacings between the Ca and
the Ti, and between the Cr and the Fe curves are in agreement with
the experimental ones. For the heavier nuclei reported in this
figure, the order between the curves is not anymore relevant
since, for Z=28,30,32 and 34, the calculated one is the same with
or without proton sub-shell closure (see fig.\ref{fig:results}).
The experimental \bu curves of the Zn and Ge isotopes cross at
N=36, one can only discuss their spacing for N$\geq$38. From N=38,
it is much more larger than the one obtained with a proton
sub-shell, but very similar to the one obtained with a proton
major-shell (long dashed curves). Between the experimental curves
of Ge and Se isotopes, the spacing is also much more large than
obtained with a proton sub-shell (see fig.\ref{fig:results}).
Above N=44, when for the Se isotopes the agreement with the
dot-dashed curves is striking, the experimental values are missing
for the Ge isotopes and no comment on the spacing between their
curves can be made. The spacing between the Zn and Ni experimental
curves depends also on N. This will be discussed in the
section~\ref{sec:Ni}. However one can note that from N=30 up to
36, the spacing between the Zn and the Ni experimental curves is
not constant, it increases from smaller up to larger than the one
obtained assuming a proton sub-shell closure (see the 20,28 and
the 30 short dashed curves). At N=40 the spacing obtained with a
major shell for the proton valence space has the good order of
magnitude as compared to the experimental one. It seems difficult
to find in the Ni isotopes
the reason of the variation with N of this spacing :\\
i) from N = 30 up to 36, the distance between the Ni experimental
\bu curve and the surrounding short-dashed and dotted calculated
curves
remains rather constant.\\
ii) from N=28 up to N=38, the experimental \bu curve of the Ni
isotopes is nearly perfectly symmetrical with respect to N=32-34,
which is the middle of a sub-shell beginning at 28 and closed at
N=38. In the next section these two points will be detailed.

In fig.\ref{fig:NpNhb}, only few experimental values are
underestimated by the calculations in the Ca and Ti isotopes. All
the \bu values of the Cr, Fe and Ni isotopes are underestimated
but moderately (not by an order of magnitude) and without shape
distortion. In fact these experimental curves seem to be simply
shifted up, the Cr and Ni ones are very regular and follow pretty
well the shapes obtained in the calculations (dotted and short
dashed lines) This will be discussed for the Ni isotopes in the
next section. For the Zn and Ge isotopes, for N$\geq$36, few
experimental values are underestimated by the calculations
performed with a major shell in protons (and sub-shell in
neutrons) but, for N$\leq$36, these calculations overestimate the
Zn and Ge experimental values. For the Se isotopes, the change is
at N$=$38~: for N$\geq$38, few experimental values are
underestimated by the calculations performed with a sub-shell in
protons and a major shell in neutrons, but below 38, these
calculations overestimates the experimental values of the light
isotopes.

The changes in the spacings and in the under-over estimations are
accompanied by the change in
 the shape of the experimental curves of Zn, Ge and Se isotopes~:
it sharpens with Z. Above N=38, the progressive shape distortion
from the nearly bell-shaped curve of the Zn isotopes, to the
deformed curve of the Ge and then to the rather peaked Se one,
heralds the very abrupt slope of the Kr, Sr and Zr experimental
\bu curves reported in fig.\ref{fig:NpNha}. For these isotopic
chains the assumption of the largest valence space for neutrons
(the 28-50 major shell), which leads to a perfect
experiment-calculations agreement for the Se isotopes with
N$\geq$44, is not sufficient to obtain an agreement in the shapes.

It results form our analysis of the figs.\ref{fig:NpNha} and
\ref{fig:NpNhb} that:
\\i) the assumption of the 20-28 shell for
valence in neutrons and in protons is in good agreement with the
experimental curves for the Ca, Ti and Cr isotopic series.
\\ii) the assumption of a sub-shell closure in the valence space of the neutrons and
of the protons is in good agreement with the experimental curves
of the Ni, and of the light isotopes of Zn.
\\iii) The combinations of valence spaces with one sub-shell closure and
one major shell (which the results are reported in long dashed and
dot-dashed lines) appear more appropriate for most of the Zn, Ge
and Se isotopes: sub-shell in neutrons and major shell in protons
for Zn (when N$\geq$36), Ge (38$\leq$N$\leq$42) and Se (up to
N=42) isotopes, then major shell in neutron and proton sub-shell
for the Ge and Se isotopes with N$\geq$42. This combination gives
also good agreement for the \bu values of the heavy (N$\geq$44)
Kr, Sr and Zr isotopes.
\\iv) the assumption of a major shell for
the valence of the protons and of the neutrons appears to be the
most appropriate combination of the valence spaces, to describe
the \bu values of the light (N$\leq$44) Kr, Sr and Zr isotopes.

In fact, several points in our analysis of the \bu curves of the
Zn, Ge, Se, Kr, Sr and Zr isotopes reported in
figs.\ref{fig:NpNha} and \ref{fig:NpNhb} lead to interpret the
content of these figures in terms of the vanishing of the protons
Z=40 and N=38 sub-shell closures.

The neutron sub-shell closure at N=38 allows to reproduce the
features of all the Ni (apart from the case of $^{68}$Ni) and Zn
isotopes. On the contrary, for the heavy Ge and Se isotopes
(N$>$42) and all the Kr, Sr and Zr ones, it has totally
disappeared.

The proton sub-shell disappearance can explain that the
contribution brought by the two protons is larger between the Ge
and Zn isotopes than between the Zn and Ni ones, as observed in
the \bu ratios curves of fig.\ref{fig:ratio}(see
sec.\ref{sec:rat}). It can also explain that the spacing between
the \bu curves of Ni isotopes and the Zn ones increases. The
distortion of the Zn and Ge experimental curves in the light
isotopes (N$\leq$36) is the illustration of the proton sub-shell
disappearance. Around N=42, the Zn, Ge, Se, Kr, Sr and Zr \bu
curves have a behavior in agreement with a proton major shell, the
proton sub-shell closure has vanished. But for N$\lessapprox$34
and N$\gtrapprox$44 these curves have a behavior in agreement with
a proton sub-shell closure assumption. Transitional curves with
distorted shapes are obtained in zones of vanishing. The closures
appear to be correlated with some of N and Z values,  calling to
mind a proton-neutron interaction effect.

To summarize, between the low \bu values and the high ones in
fig.\ref{fig:NpNha} the main phenomenon is the vanishing of the
proton sub-shell closure, between the low \bu values and the
intermediate ones in fig.\ref{fig:NpNhb} the phenomenon is the
vanishing of the proton sub-shell. Between the intermediate value
and the highest ones in fig.\ref{fig:NpNhb} there is a
transitional regime, in which some of the \bu curves have a proton
major shell and a neutron sub-shell behavior and some others
(those of the heavier nuclei) have a behavior corresponding to the
inverse combination, a neutron major shell and a proton sub-shell.
Such an inversion could traduce a p-n interaction not strong
enough in these nuclei to vanish simultaneously the both sub-shell
closures.

\begin{figure}[h]
\vspace{-0.2cm}
\includegraphics[scale=0.5]{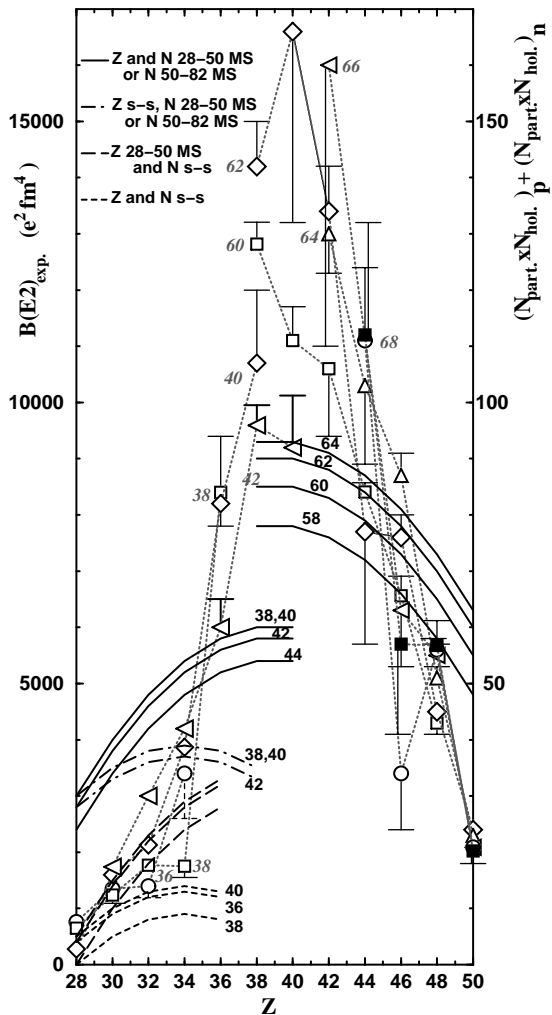}
\vspace{-0.3cm}

\caption{Joined by dotted grey lines and using different symbols,
experimental \bu values along isotonic chains for N=36 (full
circles), 38 (empty square), 40 (full grey diamonds), 42 (empty
triangles) and N=60 (full squares), 62 (empty diamonds), 64 (full
grey triangles), all taken in \cite{Raman2}. For sake of
legibility only a half of the error bars, when larger than the
symbols, are reported. In solid black lines are reported, several
of calculated curves with a major shell for each type of nucleon.
For all the curves the neutron number is explicitly given. }

\label{fig:ZpZha}
\end{figure}

\begin{figure}[h]
\vspace{-0.2cm}
\includegraphics[scale=0.5]{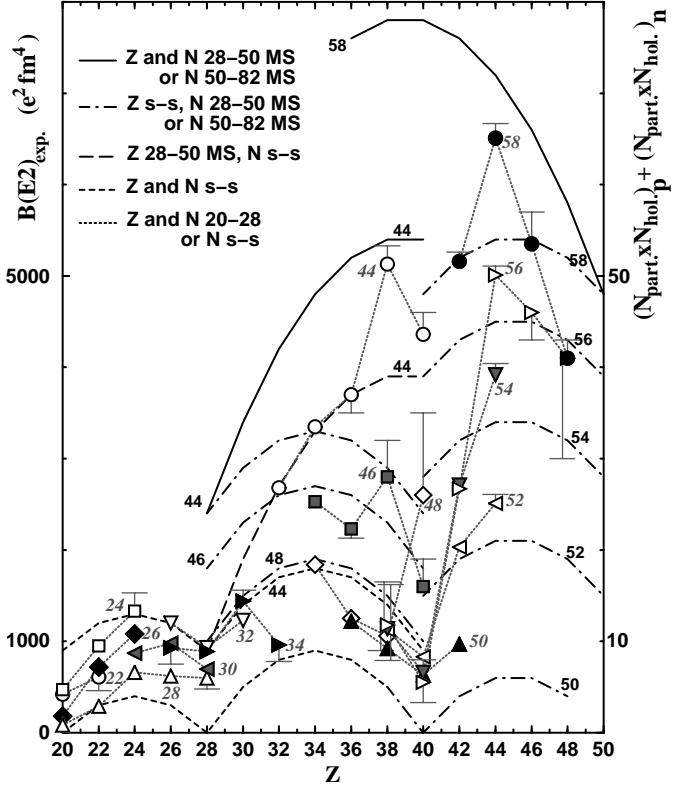}
\vspace{-0.7cm}
\caption{Joined by dotted grey lines and using different symbols
experimental \bu values along the isotonic series with neutron
number from 22 up to 36, and from 44 up to 58. All the
experimental values are from \cite{Raman2}. For sake of legibility
only a half of the error bars, when larger than the symbols, are
reported. In different black lines are reported several calculated
curves. For all the curves the neutron number is explicitly
indicated.}
\label{fig:ZpZhb}
\end{figure}

It is worth noting that the vanishing of the Z=40 sub-shell
closure has already been put forward to explain the very high \bu
values in the Zr, Mo and Ru isotopes with N$\geq$60, as compared
to the values reached for 50$\leq$N$<$60 \cite{Cas85,Ero92}. This
vanishing has been interpreted as due to the p-n interaction
\cite{Cas85,Ero92}, which was also involved in the onset of the
deformation in this region \cite{FedPitt77}. In
fig.\ref{fig:ZpZha} we present, as a function of Z
(28$\leq$Z$\leq$50), the experimental \bu curves of the N=60,62,
64 isotonic chains, which contain the Zr, Mo and Ru isotopes with
the very high \bu values discussed in \cite{Cas85,Ero92}, and the
corresponding calculated curves, assuming the 28-50 major shell
for the valence of protons and the 50-82 major shell for the
valence of the neutrons, since N is above 50. We have also
reported, in the left part of fig.\ref{fig:ZpZha}, the
experimental curves for the N=38,40 and 42 isotonic chains and the
corresponding calculated curves obtained with a 28-50 major shell
for the valence of each type of nucleon. This second group of
isotones consists in nuclei the \bu of which have been already
presented in fig.\ref{fig:NpNha} and \ref{fig:NpNhb}. The
impressive mirror symmetry with respect to Z=40 that one can
observe in fig.\ref{fig:ZpZha} between the experimental \bu curve
(slopes and widthes) of one group as compared to the other one,
confirms that a same mechanism operates in both parts of the
figure. Following the interpretation given to the highest \bu
values of the first group in \cite{Cas85,Ero92}, this mechanism is
the vanishing of a proton sub-shell closure, which then takes
place also in the second group (N=38,40, and 42) between Z=34 (Se)
and 40. Moreover, the resemblance between this fig.\ref{fig:ZpZha}
and fig.\ref{fig:ZpZhb} confirms our interpretation, in terms of
proton sub-shell closure vanishing, and the validity of the
criteria we use to draw this interpretation for the
fig.\ref{fig:ZpZhb}, in particular the one related to the
underestimation by the calculations. The similar underestimation
of the highest experimental values made, in each group of
isotones, by the corresponding calculations performed with major
shells (28-50 or 50-82 for N$>$50) for both nucleon species
confirms that the N=38,40, and 42 isotones have, above Z=34, a
28-50 major shell for the valence of each nucleon species. This is
in agreement with the scenario we can draw from the calculated
curves reported in the left part of this figure. The \bu values at
Z=28 are in agreement with the values of the curves obtained with
a neutron sub-shell closure (short and long dashed lines), below
the values obtained in the curves with a neutrons major shell.
According to the curves reported below, the proton sub-shell
closure vanishing between Z=34 and 42, is preceded by a neutron
sub-shell closure vanishing between Z=28 and 34. The neutron
sub-shell closure vanishing is not sharp, between Z=28 and 34
there is a transitional region in which the combinations of major
shell for one type of nucleon and a sub-shell closure for the
other nucleon species seem appropriate.

In fig.\ref{fig:ZpZhb}, as a function of Z (20$\leq$Z$\leq$50),
are reported the experimental \bu curves of isotonic chains for
44$\leq$N$\leq$58 and for N$\leq$36. The experimental curves
reported here as a function a Z appear in general more distorted
than the ones reported as a function of N in fig.\ref{fig:NpNhb}.
However one finds again some of the features analyzed in
fig.\ref{fig:NpNhb}~: the different orders of magnitude, the
changes in the spacing between the curves of two consecutive N
values, the presence of a minimum at the sub-shell closure,
followed by a corresponding maximum, the shapes distortion
announcing the peaked shapes of fig.\ref{fig:ZpZha}. The
similarities between the distortion of the N=44, 58, 56 curves as
a function of Z in fig.\ref{fig:ZpZhb} and the Se, Ge ones as a
function of N in the fig.\ref{fig:NpNhb} is really impressive.
Once again one can deduce that the same mechanisms operate in
fig.\ref{fig:ZpZhb} and in fig.\ref{fig:NpNhb}: firstly the
vanishing of one sub-shell closure leading from the low \bu values
obtained with two sub-shell closures to the intermediate values
corresponding to one of the two dissymmetrical combinations of
valence spaces involving one major shell and one sub-shell
closure; then an inversion of the proton and neutron valence space
size leads to the second dissymmetrical combination and to the \bu
highest values. In fig.\ref{fig:NpNhb}, the proton sub-shell
closure vanishing explains the large spacing between the curves of
the middle of the figure and in fig.\ref{fig:ZpZhb} it comes from
the neutron sub-shell closure vanishing. But, as in the
fig.\ref{fig:NpNhb}, the simultaneous vanishing of the two
sub-shells closure cannot be obtained, this is either one or the
other. And in fig.\ref{fig:ZpZhb}, as in fig.\ref{fig:NpNhb}, the
second dissymmetrical combinations of valence spaces leads to an
impressive agreement between one calculated curve and the
corresponding experimental one. It is between the N=44
experimental curve and the long dashed one (proton major shell and
neutron sub-shell closure) in fig.\ref{fig:ZpZhb}, and between the
Z=34 experimental curve and the dot-dashed one (proton sub-shell
closure and neutron major shell) in fig.\ref{fig:NpNhb}.

The interpretation in terms of the vanishing of the Z=40 sub-shell
closure put forward to explain the very high \bu values in the Zr,
Mo and Ru isotopes with N$\geq$60 \cite{Cas85,Ero92}, confirms the
conclusions we have drawn from our analysis of the different
figures. The deviation of the Zn experimental \bu curve with
respect to the Ni one, in fig.\ref{fig:NpNhb}, is explained by the
proton sub-shell vanishing. However one feature in the Ni
experimental \bu curve remains unexplained, its interpretation is
the matter of the next section.

\section{\label{sec:Ni}Interpretation of the opposite Ni and Zn \bu variations between N=38 and 40}

Let us firstly comment the increase between N=38 and 40 in the \bu
curve of the Zn isotopes. Our interpretation of this increase is
that it is, as for all the isotopic series with 30$\leq$Z$\leq$34,
due to the neutron sub-shell closure which imposes a minimum at
N=38 followed by an increase up to N=44. This interpretation is at
variance with the one given in \cite{Sor02}. The previous
explanation was an increase of the deformation between N=38 and 40
allowing the Zn isotopes to escape from $\nu 1g9/2$ at the
contrary of the Ni isotopes, spherical, in which this
orbital hinders the quadrupole excitations.\\

\begin{figure}[ht]
\vspace{-0.2cm}
\includegraphics[scale=0.5]{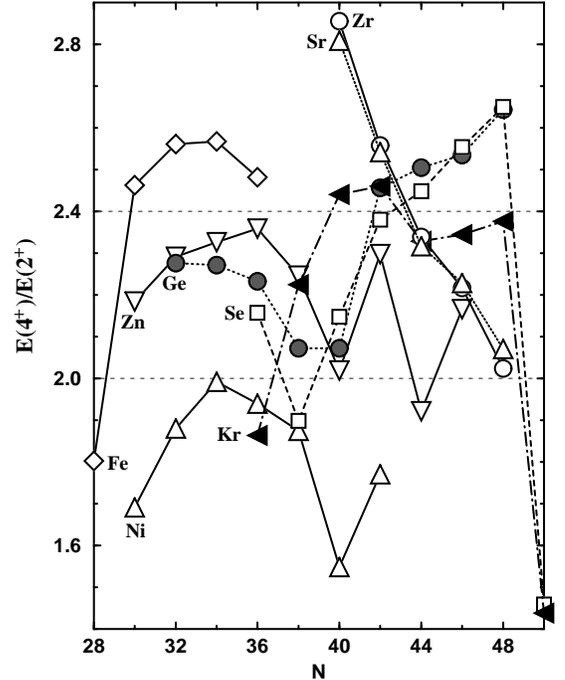}
\vspace{-0.3cm}

\caption{Joined with different lines and using different symbols,
experimental $E(4^+_1)/E(2^+_1$) values for the Fe, Ni, Zn, Ge,
Se, Kr, Sr and Zr isotopes. The energies are taken from the ENSDF
file \cite{ENSDF}. The limit given in \cite{Casten} for the near
magic nuclei, 2.0, and for the vibrators, 2.4 are reported in
dot-dashed lines.}

\label{fig:4ov2}
\end{figure}

Although bringing into play 2 more units in spin than the \bu
values, we will analyze the information provided by the
$E(4^+_1)/E(2^+_1)$ ratios. According to the ref.\cite{Casten}
this ratio allows to separate nuclei in three ranges, the magic
and near magic nuclei for $E(4^+_1)/E(2^+_1)$$<$2.0, the vibrators
between 2.0 and 2.4, the last limiting value being 3.3 for the
rotors. It gives then indications on the deformation of nuclei
nearly at the bottom of their ground state band. On
fig.\ref{fig:4ov2} are reported the $E(4^+_1)/E(2^+_1$) ratios as
a function of N for the Fe up to the Zr isotopic series. The Kr
and Sr nuclei -- for which the isotope shift measurements
previously mentioned in sec.\ref{sec:rat} indicates a large change
of deformation for N$\geq$38 -- exhibit also a large change of the
$E(4^+_1)/E(2^+_1)$ ratio, and good signs of deformation with
values above 2.4. But their $E(4^+_1)/E(2^+_1)$ evolve in
opposition whereas the isotope shift measurements give similar
evolution of their deformation. In the same way, the very low
values of the $E(4^+_1)/E(2^+_1)$ ratio in $^{72,74}$Kr are
inconsistent with the deformation parameters extracted from the
isotope shift measurements \cite{Keim}. All this is probably the
consequence of the shape coexistence phenomenon, known to strongly
perturb the low energy spectrum in the $^{72-78}$Kr
\cite{Dejb90b}. It is worth noting that, in Sr and Zr, the
$E(4^+_1)/E(2^+_1)$ ratios evolve similarly as the isotope shifts
do from N=50 up to 62 \cite{Campbell}. On fig.\ref{fig:4ov2} one
can note the parallelism of the Zn and Ni curves from N=30 up to
42, indicating that the difference of deformation stays constant
between their isotopes, in particular during the decrease of the
$E(4^+_1)/E(2^+_1)$ ratio between N=38 and 40. This appears in
contradiction with an increase of deformation, between N=38 and
40, in the Zn isotopes as compared to the Ni ones. Moreover at
N=40, end of the $E(4^+_1)/E(2^+_1)$ decrease, the Zn value is
equal to 2.0, the near magic nuclei limit. No experimental
evidence of deformation in the Zn isotopes at N=40 appears from
fig.\ref{fig:4ov2}.

Let us now come back to the particularity of the experimental \bu
curve of the Ni isotopes reported in fig.\ref{fig:NpNhb}, the
decrease between N=38 and 40. One can note that, in the Ni
isotopes, the experimental \bu value at N=38 is equal to the one
at N=28. This in agreement with a neutron sub-shell closure at
N=38 since in this case the neutron contribution is null both at
N=28 and at N=38. This \bu value, at N=40, smaller than the one
measured at N=28 and 38 is then disconcerting. This exceeds the
consequence of the simple change of parity in the neutron
single-particle level scheme as the one coming from $g_{9/2}$. At
the very most, such a parity difference can annihilate only the
odd neutron contribution such as the 1p-1h one
\cite{Grawe,Sor02,Lang03}. Therefore, even with a total neutron
contribution equals to zero, the \bu value at N=40 would be equal
to those measured at N=28 and 38, not lower. The curve of Ni
isotopes put then light on a proton effect. That can also be
deduced from the Ca experimental \bu curve. Whereas the \bu values
of the $^{40,48}$Ca isotopes are nearly equal to zero as expected
with null contributions from neutrons (N=20,28) and from protons
(Z=20), this is not the case in the $^{56,66}$Ni (N=28,38 and
Z=28) isotopes although the experimental \bu curve increases
between N=28 and 30 as expected after a neutron shell closure, and
decreases between N=36 and 38 as expected before a neutron shell
closure. It is worth noting that the experimental \bu curve of the
Ni isotopes is closer to the Z=30 calculated curve than to the
Z=28 one. Indeed (see fig.\ref{fig:NpNhb}) between N=30 and 36 the
experimental curve is merged with the Z=26 calculated one assuming
sub-shell closures for both neutrons and protons (at N=28, the
experimental values of the Ni, Cr and Fe isotopes are all merged,
near above the Z=30 calculated curve). This indicates in the Ni
isotopes an additional, with respect to our formula, and constant
(at least between N=30 and N=36) proton contribution in the \bu
values between 28 and 38, which has disappeared for N=40. This is
then all the isotopic chain which has to be taken into account in
order to understand the decrease of the Ni \bu values between N=38
and 40, which is not only an effect between the $^{56}$Ni (N=Z
nuclei) and $^{68}$Ni (N=40) isotopes has proposed in recent works
\cite{Sor02,Lang03}.

Moreover, at N=40, the experimental \bu value of the $^{68}$Ni is
in rather good agreement with the one expected from calculations
for Z=28 with two sub-shell closures (see fig.\ref{fig:NpNhb}),
the agreement resembling the one obtained for the Ca isotopic
series. If, at N=38, the additional proton contribution were null
(as it is at N=40), one would assist to an increase between a low
\bu value at N=38 and the \bu value at N=40 of the order of what
is observed in the Zn and Ge isotopes. The decrease in the Ni
isotopes between N=38 and 40 is really misleading, it dissimulates
a N=38 neutron sub-shell closure behavior. In fact the $^{68}$Ni
\bu value is in agreement with an increase following the minimum
at N=38 as in the curves obtained with a neutron sub-shell closure
at N=38.

\begin{figure}[h]
\vspace{-0.3cm}
\includegraphics[scale=0.5]{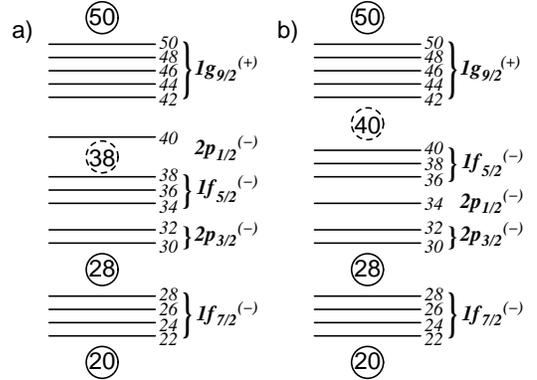}
\vspace{-0.2cm}
\caption{Order of single-particle levels for both protons and neutrons at null deformation taken a) from ref.\cite{Naz85},
b) from ref.\cite{RingSchuck}.}

\label{fig:lev}
\end{figure}

The presence of a N=38 gap constrains the order of the neutron
single-particle levels, as the Z=40 sub-shell closure constrains
the order of the proton ones. A gap can appear for particle number
equals to 38 only in the single-particle level scheme obtained in
the reference \cite{Naz85}, the order of which is reported in
fig.\ref{fig:lev}a, between the $1f_{5/2}$ and the $2p_{1/2}$
orbitals. On the contrary the order of the single-particle levels
taken from \cite{RingSchuck} and reported in fig.\ref{fig:lev}b
allows only a gap for particle number equals to 40, between the
$1f_{5/2}$ and the $1g_{9/2}$ orbitals. One will then assume in
the following for the neutrons the order of the single-particle
levels a) from ref.\cite{Naz85} and for the protons the order of
the single-particle levels b) from ref.\cite{RingSchuck}.

\begin{figure}[h]
\vspace{-0.3cm}
\hspace{-0.5cm}
\includegraphics[scale=0.5]{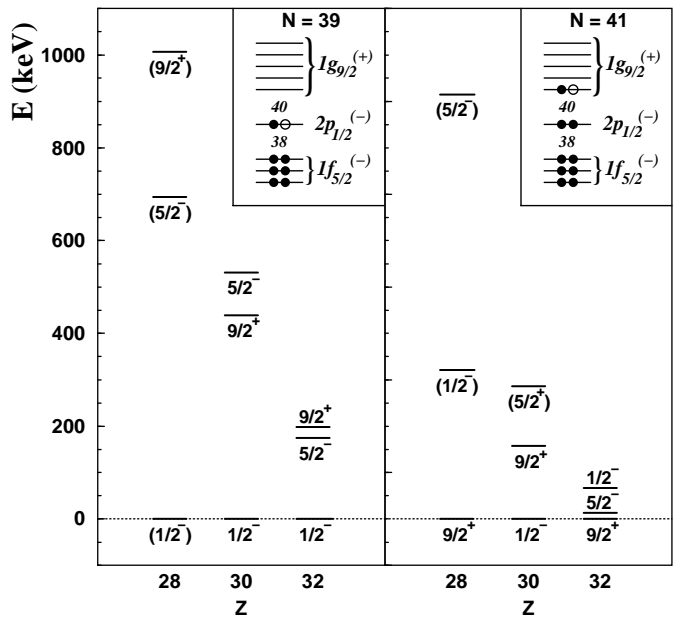}
\vspace{-0.7cm}

\caption{Experimental excitation energies of the first three states in the odd Ni, Zn
and Ge isotopes a) at N=39 and b) at N=41. In each case, the order
of the neutron single-particle levels with a schematic occupation
corresponding to the ground state of the Ni isotope is proposed.
The experimental data are taken from \cite{ENSDF}.}

\label{fig:gaps}
\end{figure}

The N=38 sub-shell closure in the Ni isotopes is confirmed by the
excitation energies of the first excited states observed in the
odd Ni isotopes with N=39 and 41 as compared to the excitations
energies in the odd N=39 and 41 Zn and Ge isotopes. The ground
state and the two first of the observed excited states are
reported in fig.\ref{fig:gaps}.

At N=39, the ground state in the Ni, Zn and Ge isotopes is a
$1/2^-$ state as expected when arising from one particle (or one
hole) in $\nu 2p_{1/2}$, the two excited states are $5/2^-$ and
$9/2^+$ as expected from a hole in $\nu 1f_{5/2}$ and a particle
in $\nu 1g_{9/2}$. At N=41, in the Ni and Ge isotopes the ground
state is a $9/2^+$ state as expected when arising from one
particle (or one hole) in $\nu 1g_{9/2}$ but in the Zn isotope it
is a $1/2^-$ state.

At N=39 and 41, in the Ni isotopes the $\nu 2p_{1/2}$ orbital is
isolated from $\nu 1f_{5/2}$ and from $\nu 1g_{9/2}$, confirming
the N=38 gap and the order of the single-particle level scheme
given in the figure. The gap at N=38 appears even stronger in the
odd Ni isotopes with N=39 than in its isotones of Zn or Ge, this
will be discussed in the next section. The spacing at N=40 between
$\nu 2p_{1/2}$ and $\nu 1g_{9/2}$ is also apparent in the Ni
isotope, illustrated by the spacing between the ($5/2^-$) and the
($9/2^+$) states. In the Zn and Ge isotopes these two excited
states are very near in energy, and both have a decreasing
excitation energy, giving a picture of the weakening of the N=38
sub-shell closure before it vanishes.

It is worth noting that the effect of the size of this energy
spacing at N=40, on the \bu value of $^{68}$Ni has been discussed
in terms of a N=40 gap counteracted by neutron pair scattering in
$^{68}$Ni \cite{Sor02} and has been recently re-discussed in the
frame of calculations involving a non null 2p-2h neutron
contribution to the \bu value in this nucleus \cite{Lang03}.
Unfortunately the first calculations have been reported only for
N$\geq$34, and in the latter calculations the \bu value calculated
at N=38 is greater than the one at N=28 and the \bu value
calculated at N=40, where the gap is assumed, is equal to the one
at N=28 \cite{Lang03}. The striking \bu value lower in $^{68}$Ni
than in $^{56}$Ni escapes to these calculations. They therefore
demonstrate that to calculate in QRPA a same \bu value at N=38 and
at N=28, a N=38 gap is needed as in our simple model calculations.
The experimental \bu curve of the Ni isotopes implies a gap at
N=38, but an energy spacing between the single-particle levels at
N=40 appears from the spectrum of the odd Ni excited states and
not from the Ge and Zn ones. It can also play a part in the
explanation of the origin and the annihilation of the additional
proton contribution at N=40 which will be discussed now.

\section{\label{sec:pn}Effects of the p-n interaction}

It has been pointed out \cite{FedPitt77,Cas81} that the p-n
interaction can be strong enough to reduce (and even to eradicate)
subshell gaps. This reduction goes with the promotion of particles
into orbitals which were above the gap and which can drive
deformation \cite{FedPitt77,Cas81,Ciz97}. The p-n interaction is
favored when there is a large overlap between the proton and the
neutron active orbital. Taking ($n_p$ $l_p$ $j_p$) to characterize
the proton orbital and ($n_n$ $l_n$ $j_n$) the neutron one, the
angular overlap is maximized when $j_p$$\approx$$j_n$$\gg$1, and
the radial overlap is maximum for $n_p$=$n_n$ and
$l_p$$\approx$$l_n$ \cite{De-Sha53}. Among the single-particle
levels reported in fig.\ref{fig:lev}a,b, the 1f$_{7/2}$,
1f$_{5/2}$ and 1g$_{9/2}$ orbitals can realize these conditions.

If, as in sec.\ref{sec:Ni}, one assumes for protons the order of
single-particle levels displayed in fig.\ref{fig:lev}b, the
$\pi$1f$_{5/2}$ orbital is occupied for Z=36, 38 and 40. These
proton numbers correspond to the Kr, Sr and Zr nuclei which
present the highest \bu values (i.e. the largest collectivity)
reported in fig.\ref{fig:NpNha}. These large \bu values are
obtained for a restricted number of their isotopes, those with
N=40$\pm$2. Assuming for neutrons, as in sec.\ref{sec:Ni}, the
order of single-particle levels reported in fig.\ref{fig:lev}a,
the neutron number N=38 corresponds to the occupation up to the
last level of $\nu$1f$_{5/2}$, N=40 to the occupation of
$\nu$2p$_{1/2}$, and N=42 to the occupation of one level of
$\nu$1g$_{9/2}$. This is nearly the situation described in
\cite{FedPitt77}, protons in $\pi$1f$_{5/2}$ and neutrons
occupying, for N$>$62, the $\nu$1g$_{7/2}$. But the polarization
effects \cite{FedPitt77} induce, from N=60 so with an anticipation
of four neutrons in these even-even nuclei, the promotion of
neutrons into this high-j orbital. In the Kr, Sr and Zr isotopes
with N=40$\pm$2, the protons occupy the $\pi$1f$_{5/2}$ orbital,
and in neutron this is the $\nu$1g$_{9/2}$ orbital which the
occupation is anticipated from N=38 when it is expected from N=42.
Neutrons are promoted and the $\nu$1g$_{9/2}$ orbital is lowered.
The eradication of any gap between $\nu$1f$_{5/2}$ and
$\nu$1g$_{9/2}$ can occur. This increases the number of levels
N$_h$ of the neutron valence space, and produce the gain of
collectivity observed for N=40$\pm$2. This phenomenon is expected
to be maximum at N=40, since at N=42, the $\nu$1g$_{9/2}$ is
naturally occupied. At N=40 the ratio of the cost in energy for
promotion over the gain in valence space is maximal. It is worth
noting that, among the high values of \bu observed in
fig.\ref{fig:ZpZha}, those of Sr, Zr and Mo isotopes, for
60$\leq$N$\leq$64, have been previously interpreted not only by
the anticipated promotion of neutrons into $\nu$1g$_{7/2}$, but
also by the promotion of protons from below the Z=40 sub-shell
closure, nearly annihilated, into $\pi$1g$_{9/2}$
\cite{FedPitt77,Cas81}. In fig.\ref{fig:ZpZha}, one can observe
that the same order of magnitude is reached by the highest \bu
values of these previous Sr, Zr and Mo isotopes (with
60$\leq$N$\leq$64) and by those of the Kr$_{36}$, Sr$_{38}$ and
Zr$_{40}$ isotopes with 38$\leq$N$\leq$42. This implies, in
agreement with our conclusions of the sec.\ref{sec:ana}, that in
these Kr$_{36}$, Sr$_{38}$ and Zr$_{40}$ isotopes, as in the Sr,
Zr and Mo ones, the proton sub-shell closure is certainly reduced
or vanished. This comes in addition to, in the Kr, Sr an Zr
isotopes, the vanishing of any neutron gap, or spacing, between
$\nu$1f$_{5/2}$ and $\nu$1g$_{9/2}$. Starting from the occupation
of $\pi$1f$_{5/2}$, the N=38 gap and the N=40 spacing are
annihilated, neutrons are promoted on $\nu$1g$_{9/2}$. And as in
\cite{FedPitt77,Cas81} the occupation of the $\nu$1g$_{9/2}$ in
the Kr, Sr, Zr isotopes allows the polarization effects to induce
the promotion of protons in $\pi$1g$_{9/2}$. In the Kr, Sr, Zr
isotopes both types of particles are promoted, as in
\cite{FedPitt77}. But here, both sub-shell gaps are vanished
thanks to the mutual polarization effect induced by the p-n
interaction.

In the Zn$_{30}$, Ge$_{32}$ and Se$_{34}$ isotopes, protons have
to be promoted from $\pi$2p$_{3/2}$ and $\pi$2p$_{1/2}$ into the
$\nu$1f$_{5/2}$ orbital, or better into the high-j $\pi$1g$_{9/2}$
one, this is possible from N=34, thanks to the occupation of the
$\nu$1f$_{5/2}$ orbital allowing the p-n interaction and the
polarization effect to take place. It appears that when the
preliminary promotion concerns protons, it weakens the p-n
interaction~: as shown in fig.\ref{fig:NpNhb}, in the Zn, Ge and
Se isotopes only the proton sub-shell closure is clearly vanished
from N=34 up to N=40. Below N=38, the only \bu value giving sign
of the neutron sub-shell vanishing is the Se one at N=36. For
neutron numbers equal to 40, the p-n interaction must be strong
enough to anticipate, as reported in \cite{FedPitt77}, the
occupation of $\nu$1g$_{9/2}$ by promoting neutrons, and to ignore
the N=40 spacing of the single-particle level scheme. Indeed, at
N=42 when the neutrons occupy ``naturally" $\nu$1g$_{9/2}$, the
mutual polarization effects between 1f$_{5/2}$ and 1g$_{9/2}$, in
the Ge and Se isotopes, are strong enough to lead to the vanishing
of the neutron sub-shell closure~: the Ge and Se \bu curves
decrease between N=42 and 44, the Se one has, afterwards, a
neutron major-shell behavior. Nevertheless, the gain in
collectivity is always lower in the Zn, Ge and Se than in the Kr,
Sr and Zr isotopes. At N=42 where the gain is maximal, the
distance between the experimental values and the relevant
calculated curve is considerably lower for the Ge and Se nuclei,
than for the Kr, Sr and Zr. On the other hand , it is worth noting
that the coupling of the pairing mode to other modes of excitation
such as the proton-neutron coupling \cite{Kumar78} has been put
forward, as well as pairing fluctuations \cite{Kumar78,Ardo78} to
explain the particularities of the Ge isotopes. The p-n
interaction can play the role of the pairing fluctuations, acting
also on the size of the spacing between single-particle levels of
one or both nucleon species.


Finally, in the Ni isotopes, the ``natural'' occupation of
$\pi$1f$_{7/2}$ make possible in all the isotopes, the
polarization effects to take place, with the promotion of
particles. As already discussed, the Ni experimental \bu curve
puts light on an additional and nearly constant (between N=30 and
36) proton contribution as compared to the calculations. Indeed
the agreement obtained by the Z=26 calculated values with a
sub-shell closure for both nucleon species is an indication for
the promotion of two protons above the Z=28 gap. It is worth
noting that in $^{56}$Ni the promotion of protons over the Z=28
gap due to the strong p-n interaction in the N=Z nuclei has been
reported in ref.\cite{Sor02}. It was also reported that, in the
$^{64-74}$Ni isotopes, 80$\%$ of the \bu values comes from the
proton core excitation into $\pi$1f$_{7/2}$.(Nevertheless in these
shell model calculations the single-particle level scheme present
only one gap at N=40, and not a N=38 gap and a spacing at N=40.)
The promotion of protons can be the effect of a mutual
polarization between the $\pi$1f$_{7/2}$, $\pi$1f$_{5/2}$ and
$\nu$1f$_{5/2}$ orbitals, into which neutrons are probably
promoted. But this promotion has no strong effect on the shape of
the experimental \bu curve. Indeed, the promotion of protons above
the Z=28 gap seems to cost a great part of the gain in
collectivity. In fig.\ref{fig:ZpZhb}, the N=28 experimental \bu
curve is, as a function of Z, flat from Z=24 up to Z=28, when it
is expected to decrease due to the proximity of the Z=28 gap. The
Z=28 gap seems to be very slightly diminished, but too large to be
vanished, there is a flattening in the curve not an increase. The
other picture of this flattening is in the fig.\ref{fig:NpNhb}, at
N=28, the near equality of the experimental \bu values of the Cr,
Fe and Ni isotopes. (The shift of the Cr and Fe experimental \bu
curves can be then also explained by the proton promotion above
the Z=28 gap and the slight weakening of it).

Therefore at N=28 in the Ni isotopes when $\nu$1f$_{7/2}$ is
filled, as at N=38 when it is $\nu$1f$_{5/2}$ which is filled, the
p-n interaction appears to be strong enough to promote some more
particles, the experimental Z=28 curve in fig.\ref{fig:NpNhb} is
slightly deformed as a function of N at these two points. One can
note the impressive symmetry of the slight deformation at these
two points. The equality of the \bu values of $^{52}$Cr, $^{54}$Fe
and $^{56}$Ni is in favor of an additional promotions of two
protons above the Z=28 gap at these two points N=28 and N=38. The
neutrons could also have here a part in the gain of collectivity,
and be promoted over the N=38 gap slightly diminished, but
maintained~: the \bu curve is still increasing from N=28 up to
N=30 and decreasing between N=36 and N=38. Nevertheless, the
neutron promotion is less probable than the proton promotion, the
decrease of the \bu value between $^{66}$Ni and $^{68}$Ni
indicates that the gap resulting from the N=38 gap followed by the
N=40 spacing is non negligible for the p-n interaction. Above
N=38, the complete filling of $\nu$1f$_{5/2}$, the next neutron
orbital favoring the p-n interaction is $\nu$1g$_{9/2}$. But the
p-n interaction appears not strong enough to continue to promote
protons from below the Z=28 gap when neutrons are as far as above
the N=38 one, or even farther if one includes the spacing at N=40
of the single-particle level scheme. This explains the loss at
N=40 of the additional proton contribution to the \bu values
observed in the Ni isotopes. No promotion of neutron to
$\nu$1g$_{9/2}$, no lowering of $\nu$1g$_{9/2}$, and no weakening
of the N=38 sub-shell closure are observed at N=40, the $^{68}$Ni
experimental \bu value is very near from the one calculated
assuming sub-shell closure for both nucleon species.

The Z=40 and N=38 sub-shell closures appear to depend strongly on
the proton and neutron high-j orbital occupation. In that sense,
as they can be vanished, they may be not considered as inducing
magicity.

\section{\label{sec:proba}Probability calculations}
\begin{figure}[h]
\vspace{-0.4cm}
\includegraphics[scale=0.5]{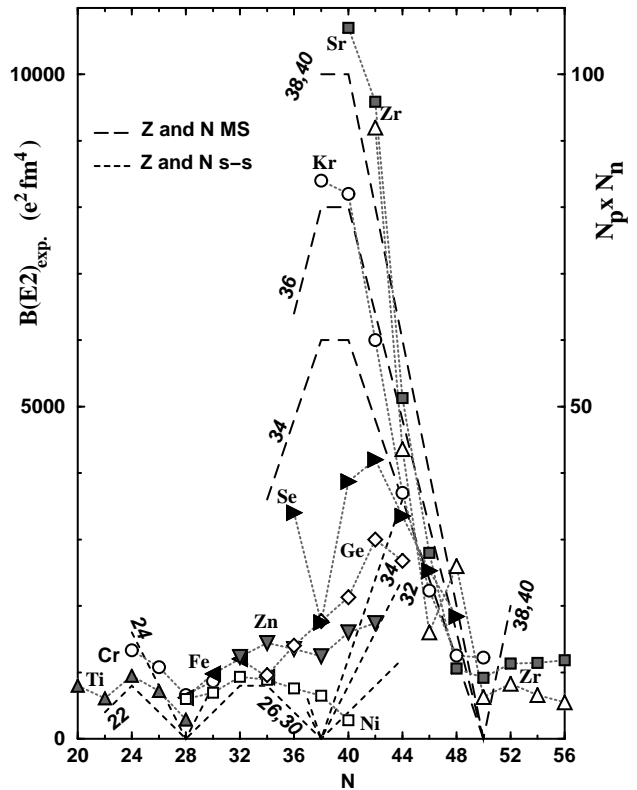}
\vspace{-0.3cm}

\caption{
Joined by dotted grey lines experimental \bu values of Ti (full
grey triangles), Cr (empty circles), Fe (full black left
triangles), Ni (empty squares), Zn (full grey triangles down), Ge
(empty diamonds), Se (full black triangles right), Kr (empty
circles), Sr (full grey squares) and Zr (empty triangles up)
isotopes. For the error bars see
fig.\ref{fig:NpNha},\ref{fig:NpNhb}. The calculated product of
number of protons and neutrons (or holes for both types of
particles past the mid-shells) N$_p$$\times$N$_n$ are reported~:
in long-dashed line for the double major shell, protons and
neutrons, in short-dashed line for both sub-shell closures (N=38
and Z=40) which is, for Z$\leq$34, the same as with the sub-shell
closure N=38 and full major shell in protons. For all the curves
the proton number is explicitly indicated. }

\label{fig:BpBn}
\end{figure}

It results from the discussion presented in the sec.\ref{sec:ana}
that the seniority formula can reproduce neither the profile nor
the magnitude of the \bu curves in case of very high collectivity.
On the other hand, the \bu value is known to depend on the product
\cite{Cas85} N$_p$N$_n$, of the number of protons (or protons
holes above the mid-shell) by the number of neutrons (or neutrons
holes above the mid-shell) in the valence shells. (This product is
also equal to four times the product N$_\pi$N$_\nu$ of the number
of the proton bosons by the number of the neutron bosons
\cite{Talmi,Cas85}.)
 As shown in fig.\ref{fig:BpBn}, when normalized to the experiment using a
scale giving 100 e$^2$fm$^4$ per unit of N$_p$N$_n$, the
calculated N$_p$N$_n$ curves present some similarities with the
experimental ones. Firstly, the relative magnitude of the Sr, Zr
and Kr, and Cr, Ti curves are very well reproduced. The agreement
is only reasonable for the Fe, Zn, Ge and Se curves. (For the Ni
isotopes, which have N$_p$=0, no comparison with the experimental
values can be drawn.) Secondly, the calculated curves for Se, Sr,
Zr and Kr converge to a unique point at the end of the shell, as
observed experimentally. However, these N$_p$N$_n$ curves present
a systematic disagreement with the experimental ones: the
experimental slopes are larger than the calculated ones. In the
well known correlation between \bu value and N$_\pi$N$_\nu$, the
latter difference plays certainly an important part in the
dispersion obtained in the \bu curve versus N$_p$N$_n$ (or
N$_\pi$N$_\nu$) of \cite{Cas85}, as well as in the curves of the
effective proton numbers N$_p$ or N$_n$ versus Z or N
\cite{ZhaoChen95}, this latter dispersion making difficult the
sub-shell closure localization \cite{ZhaoChen95}.

Another type of bell-shaped curve is obtained by considering that
the excitation probability is proportional to the number of ways
of putting P pairs of particles on L levels, i.e proportional to
$(^L_P)=\frac{L!}{P!(L-P)!}$.

Assuming a strong p-n interaction that mixes up protons and
neutrons as well as the proton and neutron high-j shells, one can
calculate the curves obtained involving a unique shell of L
levels, and on which pairs (of protons as well as of neutrons) can
be excited.

\begin{figure}[h]
\includegraphics[scale=0.5]{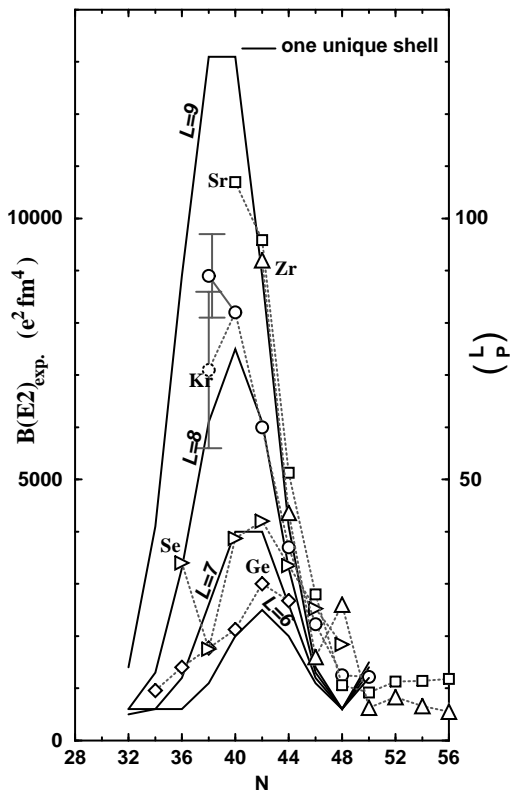}
\vspace{-0.3cm}
\caption{Joined by dotted grey lines, experimental \bu values for
Ge (empty diamonds), Se (empty triangles right) , Kr (empty
circles), Sr (empty squares) and Zr (empty triangles up) isotopes.
The calculated curves with $(^L_P)$ assuming one only shell, with
L single-particle levels, are reported in solid lines. For these
curves the level number L is explicitly indicated.}
\label{fig:fact}
\end{figure}

In fig.\ref{fig:fact} the calculated curves giving the best
agreement with the experimental \bu values of Ge up to the Zr
isotopes are reported in solid lines. The number of levels of the
unique shell fitting with the experimental \bu values is L=9 for
the Sr and Zr isotopes, L=8 for the Kr isotopes, L=7 for the Se
isotopes and, finally, L=6 for the Ge isotopes. In addition to
reasonable relative heights and slopes very near from the
experimental ones --at least nearer to the ones obtained in the
N$_p$N$_n$ calculations see fig.\ref{fig:BpBn} -- these $(^L_P)$
curves point out the alternation in the number of points which
appear at the maximum of the curves: one for Ge, two for Se. In
the case of $^{74}$Kr we have reported the results from the two
life time measurements using the recoil distance method
\cite{Tab90,Roth} which are not consistent although they are not
independent as mentioned in \cite{Tab90}. The number of points at
the maximum of the Kr curve remains uncertain, as for the Sr and
Zr curves for which the values at low neutron number are missing.
Moreover it is worth noting that a shell constituted by the
gathering of 1f$_{5/2}$ and 1g$_{9/2}$, that the p-n interaction
joins up, would have 8 levels, and 9 if one includes between them
the 2p$_{1/2}$ orbital.

\section{\label{sec:conclu}Conclusion\protect}

Using the expression obtained within the seniority scheme the \bu
features observed in the Z=28-40 with N=28-50 and 52-64 nuclei are
interpreted on the basis of both N=38 and Z=40 sub-shell
eradications due to the p-n interaction.

The low \bu value in $^{68}$Ni is explained by the N=38 sub-shell
closure and the Z=28 gap, which prevent, for N above 38, the p-n
interaction to act as it does in the lighter Ni isotopes, or in
the heavier nuclei, from the Zn, Ge and Se up to the Zr isotopes.
It would be interesting to measure the $^{70}$Ni \bu value.
Indeed, the \bu value in $^{68}$Ni, as in $^{70}$Zn, indicates a
behavior following the one expected with N=38 sub-shell closure.
On another hand from N=42, the neutrons in $\nu$1g$_{9/2}$ could,
with the protons in $\pi$1f$_{7/2}$, start the p-n interaction. In
this case one could expect a proton contribution to appear as
observed between N=28 and 38. A distortion of the \bu curves as
observed in the Ge and Se isotopes could also appear, although
improbable. Indeed, even at N=42 when neutrons are occupying
$\nu$1g$_{9/2}$, there is no distortion in the case of the Zn
isotopes, due to the cost of promoting protons from
$\pi$2p$_{3/2}$ into $\pi$1f$_{5/2}$, and, in the Ni isotopes, the
Z=28 gap exhausts a large part of the gain in collectivity in the
light isotopes. The distortion, due to the p-n interaction, of the
Ge, Se curves and of the Kr, Sr and Zr isotope ones, is amazingly
reproduced by the curves of the probabilities to excite P pairs of
particles into the L levels of a unique valence shell.

\begin{acknowledgments}
The authors want to thank R. Lombard for stimulating discussions.
\end{acknowledgments}

\bibliography{myfile}

\end{document}